\documentclass[preprint, 12pt]{elsarticle}

\usepackage{amssymb}

\usepackage{amsmath,pstricks,pst-3d,multido}
\usepackage[]{graphics}
\usepackage{bbold}
\usepackage{subfigure}

\def\ket#1{\mathinner{|{#1}\rangle}}

\def\nn{\nonumber}

\journal{Annals of Physics}

\begin{document}

\begin{frontmatter}



\title{Anyonic order parameters for discrete gauge theories on the lattice}


\author[itfa,sfi]{F.A. Bais}
\ead{f.a.bais@uva.nl}
\author[itfa]{J.C. Romers}
\ead{j.c.romers@uva.nl}
\address[itfa]{Institute for Theoretical Physics, University of Amsterdam, Valckenierstraat 65, 1018XE Amsterdam}
\address[sfi]{Santa Fe Institute, Santa Fe, NM 87501, USA}

\begin{abstract}
\noindent We present a new family of gauge invariant non-local order parameters $\Delta^A_\alpha$ for (non-abelian) discrete gauge theories on a Euclidean lattice, which are in one-to-one correspondence with the excitation spectrum that follows from the representation theory of the quantum double $D(H)$ of the finite group $H$. These combine magnetic flux-sector labeled by a conjugacy class with an electric representation of the centralizer subgroup that commutes with the flux.
In particular cases like the trivial class for magnetic flux, or the trivial irrep for electric charge, these order parameters reduce to the familiar Wilson and the 't Hooft operators respectively. It is pointed out that these novel operators are crucial for probing the phase structure of a class of discrete lattice models we define, using Monte Carlo simulations.
\end{abstract}

\end{frontmatter}


\section{Discrete gauge theories and their excitations}
In two-dimensional gauge theories we distinguish two classes of particle-like excitations, which we call electric and magnetic. Electric excitations are labeled by  nontrivial representations of the gauge group, and are either put into the theory as external charges or as  dynamical fields in the action.
The magnetic excitations are labeled by topological quantum numbers, on the classical level related to solitonic sectors of the gauge theory.  In planar physics these are the magnetic fluxes and non-abelian generalizations thereof. On the quantum level these manifest themselves as particle-like excitations carrying these topological quantum numbers. 

The magnetic excitations may also carry electric charges, which leads to exotic particles called {\em dyons}. In the two dimensional setting these particle types may behave like {\em anyons} with fractional spin obeying  highly nontrivial \textit{braid statistics}.  Generally speaking, quantum groups  provide the language in which the quantum physics of two-dimensional anyonic systems is optimally casted and understood. In that sense a complete classification of all sectors of a discrete gauge theory (DGT) can be achieved within the mathematical framework of the {\em quantum double} $D(H)$  \cite{Drinfeld} of the discrete gauge group $H$ \cite{Bais:1991pe}. 

Experimental realizations  of discrete gauge theories are still modest, though there are interesting proposals for implementing them in Josephson junction networks \cite{doucot04, Doucot05a, Doucot05c} and in systems of polar molecules in optical lattices
\cite{jiang2007}. Globally symmetric implementations were explored in  \cite{bais:2006st, bais:2007bq}. In a larger context it should be pointed out that DGT's play a crucial role  in Kitaev's \cite{Kitaev:1997wr}  seminal paper on topological quantum computation, as they constitue the most elementary examples of Topological Quantum Field Theories. This simplicity follows from the fact that there is no underlying Conformal Field Theory involved as is the case for Chern-Simons theories.

\subsection{Transformations on DGT states: quantum symmetry}
We will not give a detailed account on the emergence of quantum group symmetry in DGT, this can be found in the literature \cite{deWildPropitius:1995hk}, but do present a short summary of the basics to fix the notation and introduce some key concepts required later on.

Consider the following operators acting on states in the Hilbert space of a DGT. First there is the {\em flux projection} operator, denoted by $P_h$, which acts as follows on a state $\ket{\psi}$:
\begin{displaymath}
	P_h \, \ket{\psi}=\left\{
	\begin{array}{ll}
		\ket{\psi} & \mbox{if the state $\ket{\psi}$ contains flux }h\\
		0 & \mbox{otherwise }
	\end{array}\right.
	.
\end{displaymath}
Secondly, we have the operator $g$, for each group element $g \in H$, which realizes a global gauge transformation by the element $g$:
\begin{equation}
g \, \ket{\psi} = \ket{^g\psi} , \nn
\end{equation}
where it should be noted that we have not yet modded out by the gauge group to obtain the physical Hilbert space.

These operators do not commute, and realize the following algebra:
\begin{equation}
\begin{split}
P_h P_{h'} &= \delta_{h,h'} P_h \\
g P_h &= P_{ghg^{-1}} g \, .
\end{split}
\end{equation}
The set of combined flux projections and gauge transformations $\{ P_h g \}_{h,g \in H} $ generates the quantum double $D(H)$, which is a particular type of algebra called a Hopf algebra. 
In the following we will see that its representations correspond one to one with the electric, magnetic and dyonic sectors of the discrete gauge theory.

\subsection{\label{reptheo}Representation theory of the quantum double: particle spectrum}
The representation theory of the quantum double $D(H)$ of a finite group $H$ was first worked out in \cite{Roche} but here we follow the discussion presented in \cite{deWildPropitius:1995hk} and follow the conventions of that paper.

Let $A$ be a conjugacy class in $H$. We will label the elements within $A$ as
\begin{equation}
\left\{ ^A\!h_1 , ^A\!\!h_2, \cdots, ^A\!\!h_k \right\} \in A\, ,
\end{equation}
for a class $A$ of order $k$. In general, the centralizers for the different group elements within a conjugacy class are different, but they are isomorphic to one another. Let $^A\!N \subset H$ be the centralizer for the first group element in the conjugacy class $A$, $^A\! h_1$.

The set $^A\!X$ relates the different group elements within a conjugacy class to the first:
\begin{equation}
^A\!X = \left\{ ^A\!x_{h_1}, ^A\!\!x_{h_2}, \cdots, ^A\!\!x_{h_k} \, \middle\vert \, ^A\!h_i \,=\, ^A\!x_{h_i} \, ^A\!h_1 \, ^A\!x_{h_i}^{-1} \right\}\, .
\end{equation}
This still leaves a lot of freedom, but we fix our convention such that $^A x_{h_1} = e$, the group identity element. The centralizer $^A\!N$, being a group, will have different irreps, which we label by $\alpha$. The vector space for a representation $\alpha$ is spanned by a basis $^\alpha v_j$. The total Hilbert space that combines magnetic and electric degrees of freedom, $V^A_\alpha$, is then spanned by the set of vectors
\begin{equation}
\left\{ \ket{^A\!h_i , ^\alpha\!v_j} \right\} \, ,
\end{equation}
where $i$ runs over the elements of the conjugacy class, $i = 1, 2, \cdots, \mathrm{dim}\;A$ and $j$ runs over the basis vectors of the carrier space of $\alpha$, $j=1,2,\cdots,\mathrm{dim}\;\alpha$.

To see that this basis is a natural one to act on with our flux measurements and gauge transformations, consider an irreducible representation $\Pi^A_\alpha$ of some combined projection and gauge transformation $P_h g$:
\begin{equation}
\label{eqn.gaugetr}
\begin{split}
\Pi^A_\alpha ( & P_h g ) \ket{^A\!h_i , ^\alpha\!v_j}  \\ 
&= \delta _{h, g\,^A\!h_i\,g^{-1} } \ket{g\,^A\!h_i\,g^{-1} , \sum_m D_\alpha (\tilde{g})_{mj}\,^\alpha v_m}\, ,
\end{split}
\end{equation}
where the element $\tilde{g}$ is the part of the gauge transformation $g$ that commutes with the flux $^A\!h_1$, defined as
\begin{equation}
\label{eqn.centr}
\tilde{g} \left.=\right. ^A\!\!x_{gh_ig^{-1}}^{-1}\,g\,^A\!x_{h_i} \, .
\end{equation}
This indeed commutes with the element $^A\!h_1$:
\begin{align*}
^A\!h_1\,\tilde{g} &= ^A\!\!h_1\,^A\!x_{gh_ig^{-1}}^{-1}\,g\,^A\!x_{h_i} \\
&= ^A\!\!x_{gh_ig^{-1}}\,g\,^A\!x_{h_i}\,^A\!h_1\,^A\!x_{h_i}^{-1}\,g^{-1}\,g\,^A\!x_{h_i} \\
&= ^A\!\!x_{gh_ig^{-1}}^{-1}\,g\,^A\!x_{h_i}\,^A\!h_1 = \tilde{g}\,^A\!h_1
\end{align*}
Note that when performing a successive series of gauge transformations (\ref{eqn.gaugetr}), the definition (\ref{eqn.centr}) of the centralizer element working on the charge part ensures that the left $^A\!x$ element of the first transformation equals the right $^A\!x$ element of the second, so that one is left with terms like
\begin{equation}
\label{eqn.centrmatmul}
\begin{split}
\sum_l D_\alpha & \left( ^A \! x^{-1}_{(gg')h_i(gg')^{-1}}\,g'\, ^A  \! x_{gh_ig} \right)_{ml} D_\alpha \left( ^A\! x^{-1}_{gh_ig} \,g\, ^A\! x_{h_i} \right)_{lj} \\
& = D_\alpha \left( ^A\! x^{-1}_{(gg')h_i(gg')^{-1}} \,(g' g)\, ^A\! x_{h_i} \right)_{mj},
\end{split}
\end{equation}
which is a property required later on in the definition of the lattice order parameter.
\section{The lattice formulation of a DGT}
We will first briefly introduce the spacetime lattice approach to gauge theories. This approach to gauge field theories was introduced in the context of quark confinement \cite{Wilson:1974sk}, but some research concerning discrete gauge groups \cite{Alford:1990fc,Lo:1995jr} has been conducted too. We note however that the latter papers were unable to capture the full richness of the quantum double spectrum. In particular, the dyonic sector of the theory was not touched upon.

For a more exhaustive introduction to lattice gauge theory, consider the review articles \cite{Balian:1974ts,Balian:1974ir,Balian:1974xw} or textbooks \cite{Itzykson:1989sx,Smit:2002ug}.

After setting the stage we introduce a new class of order parameters corresponding to any of the sectors in the quantum double representation theory.
\subsection{Lattices, gauge fields and actions}
We discretize spacetime into a set of sites $i,\, j,\, \cdots$ using a rectangular lattice. The gauge field $U_{ij}$, which takes values in the gauge group $H$, lives on the links $ij, \, jk, \, \cdots $ connecting sets of neighboring sites. The links are oriented in the sense that $U_{ij} = U^{-1}_{ji}$.

We note that the gauge field $U_{ij}$ takes care of the parallel transport from site $i$ to site $j$, were we to have included a matter field charged under the gauge group. An ordered product of links along a closed loop measures the holonomy of the gauge connection.

Gauge transformations are labeled by a group element $g_i \in H$ and are performed at the sites of the lattice. The gauge field transforms as
\begin{equation}
U_{ij} \mapsto g_i \, U_{ij} \, g^{-1}_j\, .
\end{equation}

The standard form for the lattice gauge field action makes use of the ordered product of links around the smallest closed loop possible, the plaquette $ijkl$:
\begin{equation}
U_p = U_{ijkl} = U_{ij}\,U_{jk}\,U_{kl}\,U_{li} \, ,
\end{equation}
which transforms as a conjugation under gauge transformations
\begin{equation}
U_p \mapsto g_i\,U_p\,g^{-1}_i\, .
\end{equation}

The action per plaquette, which corrsponds to the well-known form $F_{\mu\nu}^2$ in the continuum limit for $H=SU_N$ \cite{Wilson:1974sk}, is given by
\begin{equation}
S = - \sum_\alpha \beta_\alpha \chi_\alpha \left(U_p \right) \, ,
\end{equation}
where $\chi_\alpha$ is the group character in irrep $\alpha$ and $\beta_\alpha$ is inversely proportional to the coupling constant for irrep $\alpha$. For $SU_N$ gauge theories one usually only includes the fundamental representation and is thus left with only one coupling constant. However, we will not make this restriction here. Gauge invariance of the action is ensured by the fact the characters are class functions, in other words we consider actions where the number of independent couplings equals the number of classes i.e. the number of irreps.

With these definitions, the (imaginary time) path integral becomes
\begin{equation}
\label{eqn.pathint}
Z = \int DU \, e^{- \sum_p S(U_p)} = \prod_{ij} \frac{1}{|H|} \sum_{U_{ij} \in H} \prod_p e^{-S(U_p)}\, .
\end{equation}
\subsection{The order parameters $\Delta^A_\alpha$}
In gauge theories, the order parameters are non-local operators in one-to-one correspondence with the different particle species. Gauge invariance of these operators is a fundamental requirement, and furthermore Elitzur's theorem \cite{Elitzur:1975im} guarantees that the quantum expectation value of a non-gauge invariant operator vanishes.

Here we present a class of order parameters $\Delta^A_\alpha$ labeled by a class $A$ and centralizer irrep $\alpha$ that have a one-to-one correspondence with the particle species in the spectrum of the quantum double representation theory.

\begin{figure}

\centering

\subfigure[Construction of the operator $\Delta^A_\alpha$.]{

\begin{pspicture}(-3,-3)(3,3)
	\psset{unit=.15cm}
	\psset{viewpoint=0.3 1 0.5}
	\multido{\iter=12+12}{2}{\ThreeDput[normal=0 0 1](0,0,\iter){
		\psframe[linecolor=black,linewidth=1.5pt](12,12)}}
	\ThreeDput[normal=0 1 0](12,0,0){
		\psline[linecolor=black,linewidth=1.5pt](0,24)(0,0)}
	\ThreeDput[normal=0 1 0](12,0,0){\psdots[linecolor=black,dotstyle=*,dotscale=1](0,0)}
	\ThreeDput[normal=0 1 0](12,0,12){\psdots[linecolor=black,dotstyle=*,dotscale=1](0,0)}
	\ThreeDput[normal=0 1 0](12,0,24){\psdots[linecolor=black,dotstyle=*,dotscale=1](0,0)}
	\rput(-13,-2){$i_0$}
	\rput(-13,9){$i_1$}
	\rput(-13,20){$i_2$}
	\rput(-14,4){$U_{01}$}
	\rput(-14,15){$U_{12}$}
	\rput(4,9){$U_{p_1}$}
	\rput(4,20){$U_{p_2}$}
\end{pspicture}
\label{fig.smalldyonloop}
}
\subfigure[Location of the operator $\Delta^A_\alpha$: set of plaquettes $\Xi$ corresponding to a closed loop on the dual lattice, with  the associated loop on the real lattice. Also shown is the site $i_0$ acting as the basepoint.]{
\begin{pspicture}(-3.5,-3.5)(3,3)
	\psset{unit=.05cm}
	\psset{viewpoint=0.3 1 0.5}
	\multido{\iter=-36+12}{8}{\ThreeDput[normal=1 0 0](\iter,0,-48){
		\psframe[linecolor=gray](12,12)}}
	\multido{\iter=-36+12}{8}{\ThreeDput[normal=0 0 1](-48,0,\iter){
		\psframe[linecolor=gray](12,12)}}
	\multido{\iter=-36+12}{8}{\ThreeDput[normal=1 0 0](\iter,0,48){
		\psframe[linecolor=gray](12,12)}}
	\multido{\iter=-36+12}{8}{\ThreeDput[normal=0 0 1](48,0,\iter){
		\psframe[linecolor=gray](12,12)}}
	\ThreeDput[normal=0 1 0](60,0,-36){
		\psframe[linecolor=black](108,84)}
	\ThreeDput[normal=0 1 0](60,0,-36){\psdots[linecolor=black,dotstyle=*,dotscale=1](0,0)}
	\rput(-62,-44){$i_0$}
\end{pspicture}
\label{fig.dyonloop}
}

\caption{Two building blocks of the operator $\Delta^A_\alpha$ and the full operator.}

\end{figure}
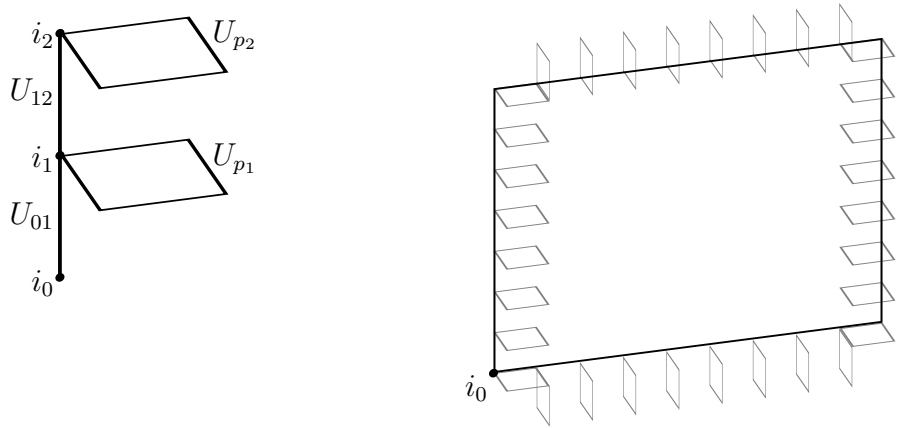
 
\subsubsection{Construction}
We will first show what the order parameter locally looks like and after that give a definition of the operator as a whole. It is well-known that in the lattice picture a magnetic flux corresponds to a nontrivial group element living on a plaquette. On the other hand, an electric charge consists of a lattice site that lies on a closed loop of links, of which the traced product is evaluated in the path integral \cite{Smit:2002ug}. In this paper we  construct a loop operator combining both flux and charge, taking proper care of the intricate interlinkage between them (\ref{eqn.gaugetr}) we alluded to before.

Consider Figure \ref{fig.smalldyonloop}. We have marked the site $i_0$, which will act as a reference point or {\em flux bureau of standards} \cite{Lo:1993hp}. The sites $i_0$ to $i_2$ are where the charge lives. Two links are named $U_{01}$ and $U_{12}$ and the two plaquette products are $U_{p_1}$ to $U_{p_2}$. We take the plaquette product to start at the point where the vertical links touch the plaquette and multiply the edges of the plaquette clockwise. Note we have left out the plaquette at the bottom; our building block for the operator will be the link pointing upward followed by the plaquette.

We now want to insert fluxes labeled by a class $A$ into the plaquettes $p_1$ and $p_2$ and make sure that a charge $\alpha$ will live at the sites $i_0$, $i_1$ and $i_2$. To insert a flux, we have to 'twist' the path integral. What we mean by this is that we want to make sure that in the integration over field configurations the dominant contribution will not come from the configuration where all the plaquettes take the trivial value in the group, but from the one where all plaquettes are trivial except the chosen plaquettes $p_1$ and $p_2$ (note that we will need to extend this set of plaquettes to one corresponding to a closed loop on the lattice; the present discussion serves as an intermediate step to arrive at the final definition.)

Let us pick a flux state $h$ in the conjugacy class $A$. We will compare fluxes at the point $i_0$. This means that the holonomy around a plaquette $p_i$ with basepoint $i_0$ will equal $h$ for all plaquettes $p_i$. The flux to be inserted at plaquette $p_1$ is thus $U^{-1}_{01}\,h\,U_{01}$, whereas the flux to be inserted at $p_2$ equals $U^{-1}_{12}\,U^{-1}_{01}\,h\,U_{01}\,U_{12}$.

We will now start at $i_0$ in Figure \ref{fig.smalldyonloop} and work our way upwards. We associate the link $U_{01}$ with the plaquette $p_1$. This association is necessary because we require a flux state to map an element in the full group $H$ to the centralizer $^AN$. The representation matrix to be put on the first link is
\begin{equation}
\label{eq.firstlink}
D ( x^{-1}_{h} \, U_{01} \, x_{U^{-1}_{01}hU_{01}} )\, ,
\end{equation}
where we have used equation (\ref{eqn.centr}) with $U_{01}$ playing the role of $\tilde{g}$, to map elements of the full group to the centralizer.

Next is the insertion of flux at $p_1$. We want the path integral to favor a configuration with a nontrivial group element at this plaquette instead of the trivial one. The action favors the trivial element, so by changing the $U_{p_1}$ appearing in the action to $\left(U^{-1}_{01}\,h\,U_{01}\right)^{-1}\,U_{p_1}$, the dominant contribution to the path integral will come from configurations where $U_{p_1} = U^{-1}_{01}\,h\,U_{01}$. The relevant operator expression to put in the path integral is thus
\begin{equation}
\label{boltztwist}
e^{S(U_{p_1})-S(U^{-1}_{01}\,h^{-1}\,U_{01}\,U_{p_1})}\, ,
\end{equation}
where the first term eats the original Boltzmann factor appearing in the path integral and is replaced by the second.

Up to the next representation matrix. This one is associated with the plaquette $p_2$, where we insert flux $U^{-1}_{12}\,U^{-1}_{01}\,h\,U_{01}\,U_{12}$. This gives a matrix
\begin{equation}
\label{eq.secondlink}
D ( x^{-1}_{U^{-1}_{01}hU_{01}} \, U_{12} \, x_{U^{-1}_{12}U^{-1}_{01}hU_{01}U_{12}} )\, .
\end{equation}
Note that the left element between brackets of this expression equals the inverse of the right element between brackets in equation (\ref{eq.firstlink}). This is crucial since it will allow us, when we make use of the representation property of these matrices, to keep gauge invariance.

The flux insertion at plaquette $p_2$ is completely analogous to the former case. This leaves us with with the following expression for (a small part of) the order parameter:
\begin{multline}
D ( x^{-1}_{h}  U_{01} x_{U^{-1}_{01}hU_{01}} ) e^{S(U_{p_1})-S(U^{-1}_{01} h^{-1} U_{01} U_{p_1})} \cdots\\
\cdots D ( x^{-1}_{U^{-1}_{01}hU_{01}} U_{12} x_{U^{-1}_{12}U^{-1}_{01}hU_{01}U_{12}} ) e^{S(U_{p_1})-S(U^{-1}_{12}U^{-1}_{01} h^{-1} U_{01}U_{12} U_{p_1})}\, .
\end{multline}
This makes clear the idea of the construction. Now we will introduce some extra notation to make the general expression not too convoluted. The different fluxes in a conjugacy class $A$ are called $h_i$. The correct group element to be inserted into the action at plaquette $p_j$ is denoted by $h^{p_j}_i$. For example, in equation (\ref{boltztwist}), this group element would be $U^{-1}_{01}\,h^{-1}\,U_{01}\,U_{p_1}$.


If we now draw a closed loop on the dual lattice, this loop pierces a set of plaquettes. We call this set $\Xi$. With this notation and the above considerations, the anyonic operator $\Delta^A_\alpha$ is given by:

\begin{equation}
\label{eqn.dyonop}
\Delta^A_\alpha (\Xi) =  \frac{1}{|A|} \sum_{h_i \in A}\prod_{j \in \Xi} D_\alpha\left( x^{-1}_{U_{j-1,j}h_i^{p_j}U^{-1}_{j,j-1}} \; U_{j-1,j} \; x_{h_i^{p_j}} \right) e^{S(U_{p_j}) - S(h_i^{p_j}U_{p_j})} .
\end{equation}
The sum over fluxes within a class is required for gauge invariance, as we will show next.
\subsubsection{Gauge invariance}
Gauge invariance of expression (\ref{eqn.dyonop}) is assured by two of its properties. First, every building block contains a product of group elements taking care of the parallel transport to a {\em single} reference point $i_0$. This already ensures gauge invariance with respect to all local gauge transformations, {\em except} at the point $i_0$.

Gauge invariance at $i_0$ is guaranteed by summing over the group elements within the class $A$. A gauge transformation at this point by an element $g$ changes the inserted flux from $h_i$ to $gh_ig^{-1}$, which corresponds to a reordering in the sum over elements in $A$.

One might argue that singling out a reference point $i_0$ seems artificial. However, the same argument that we just gave can be used to show that the point we pick is arbitrary. Selecting a neighboring point $i'_0$ would give a conjugation of $h_i$ by the element $U_{i_0i'_0}$, which would result in a reordering of the terms in the sum over elements in $A$.

\subsubsection{Closing the loop}
We built the order parameter by considering a building block of a single link and a plaquette. The full order parameter is a closed loop of these building blocks, in the sense that the links form a closed loop on the lattice and the plaquettes are pierced by a closed loops of links on the dual lattice.

To go around a corner, it might be necessary to include two links or no links at all in a single building block. This is straightforward. However, how to close the loop on itself might be less trivial at first sight. 

Since each $D_\alpha$ in equation (\ref{eqn.dyonop}) is a representation matrix of the centralizer group, we can use the property $D_\alpha(g)D_\alpha(h)=D_\alpha(gh)$. This results in  all of the $x_{h_i}$ factors canceling in the definition of the operator, except for the first and the last. However, since total group element looks like
$x^{-1}_h \, U_{\mathrm{loop}} \, x_{U^{-1}_{\mathrm{loop}}hU_{\mathrm{loop}}}$,
it is still of the form (\ref{eqn.centr}) and is thus an element of the centralizer group.

\subsubsection{Topological spin}
The dyonic sectors in a DGT posess a feature called {\em topological spin} \cite{deWildPropitius:1995hk}. This means that under a $2\pi$ rotation, their wavefunction obtains a nontrivial spin factor (for ordinary bosons or fermions this factor would be $+1$ or $-1$ respectively, but in a topological theory such as a DGT more general complex phases are allowed.) For a dyon carrying flux $A$ and charge $\alpha$, the topological spin is defined as
\begin{equation}
e^{2 \pi i s_{(A, \alpha)}} = \frac{\chi_\alpha(A)}{d_\alpha}\, ,
\end{equation}
where $d_\alpha$ is the dimension of the irrep $\alpha$.

The topological spin appears when the electric part of the loop on which the operator is evaluated winds around a single plaquette once. Let us call the loop with such a winding $\Xi'$. The same loop without the winding is called $\Xi$. The topological spin appears as
\begin{equation}
\langle D^A_\alpha (\Xi') \rangle = e^{2\pi i s_{(A, \alpha)}} \langle D^A_\alpha (\Xi)\rangle \, ,
\end{equation}
since the electric charge $\alpha$ picks up a flux in class $A$ in this operator expectation value.
\subsubsection{Comparison with known order parameters}
The order parameters that so far have been used in lattice gauge theories are the Wilson loop \cite{Wilson:1974sk} and, to lesser extent, the 't Hooft loop \cite{tHooft:1977hy}. The former corresponds to the creation and later annihilation of a particle-antiparticle pair, whereas the latter does the same for a flux-antiflux pair. The order parameters proposed here are not simple products of these two operators, but one can 'disentangle' the flux and charge part by making special choices for either the conjugacy class $A$ or the irrep $\alpha$.

When we select for the class the trivial class in equation (\ref{eqn.dyonop}), the expression reduces to the expression for the Wilson loop $W_\alpha$:
\begin{equation}
D^e_\alpha = W_\alpha = \chi_\alpha \left(U_{i_1i_2}U_{i_2i_3} \cdots U_{i_Ni_1} \right) \, ,
\end{equation}
where it should be noted that in the case of the trivial class (or any class defined by a central group element) the centralizer is the whole group, and the set of allowable centralizer representations is given by the full set of representations of $H$.

On the other hand, when a trivial group representation is selected in equation (\ref{eqn.dyonop}), the expression reduces to
\begin{equation}
D^A_1 = H^A = \frac{1}{|A|} \sum_{h_i\in A} \, \prod_{p_j\in\Xi} \, e^{S(U_{p_j}) -S(h_i^{p_j}U_{p_j})}\, ,
\end{equation}
which is an 't Hooft loop for class $A$. Our definition here is slightly different from the one appearing in the literature \cite{Alford:1990fc}, where the sum over the class is performed independently for each plaquette. The authors in that reference perform these sums independently to symmetrize over the different possible places in the plaquette where the flux insertion can take place, since {\em e.g.}
\[
h\,U_{12}\,U_{23}\,U_{34}\,U_{41} \neq U_{12}\,h\,U_{23}\,U_{34}\,U_{41} \quad \mathrm{etc.}\nn
\]
Now, since we provide a framing,  the flux insertions are all performed at the point where the electric loop touches the different plaquettes. One could achieve the same effect with definition (\ref{eqn.dyonop}), by performing a number of different twists at each plaquette (loop $\Omega$ around the plaquettes in $\Xi$) and then picking the trivial centralizer irrep. However, since only dyons have a nontrivial topological spin \cite{deWildPropitius:1995hk}, this action is immaterial.

\subsubsection{Computational complexity}
The usefulness of formulating this order parameter in the context of Euclidean lattice gauge theory (instead of {\em e.g.} using the Hamiltonian approach \cite{Kitaev:1997wr}) lies in the fact one can use Monte Carlo (MC) methods to study its behavior and probe the phase structure in the parameter space of the theory.

As equation (\ref{eqn.dyonop}) does look quite complex one might fear that the operators we have proposed are not very convenient to use in  MC simulations. We like to argue out that this is not the case and will further substantiate this by explicit calculations in a future publication.
We first like to point out that  in MC simulations most CPU time is by far used for generating new configurations. The time it takes to perform a measurement of an operator like (\ref{eqn.dyonop}) within a single configuration is negligible compared to that of generating a new field configuration. 
The second point is concerned with obtaining a good sampling resolution on the generated MC configurations. The problem here is that the dominant contribution to the expectation value of the operator we propose will not coincide with the dominating configurations generated by the MC algorithm in the absence of the operator. This problem has been studied in the context of $U_1$ gauge theory and magnetic monopole operators \cite{D'Elia:2006vg}, and their results are directly applicable here.

\section{Conclusion and outlook}
We have proposed a family of order parameters $\Delta^A_\alpha$ for discrete gauge theories on a Euclidean lattice. They are in one-to-one correspondence with the excitation spectrum that follows from the representation theory of the quantum double $D(H)$ of the finite group $H$. Since they explicitly contain the magnetic flux label $A$ and electric charge label $\alpha$, they directly allow quantum double physics to be studied on the lattice. In particular cases these order parameters reduce to the familiar Wilson and the 't Hooft loop operators respectively. 

These order parameters can be studied using Monte Carlo simulations. In a forthcoming paper, we will report on the phase diagram of a non-abelian discrete gauge theory using the new order parameters proposed here.

The authors would like to thank Jan Smit and Joost Slingerland  for  illuminating discussions. JCR is supported by the Stichting voor Fundamenteel Onderzoek der Materie (FOM) of the Netherlands.



\bibliography{biblio.bib}

\end{document}